\documentclass[aps,prd,onecolumn,groupedaddress,showpacs,nofootinbib,amssymb
]{revtex4}
\usepackage[dvips]{graphicx}
\usepackage{amssymb}
\usepackage{amsmath}
\usepackage{graphicx}
\usepackage{amsfonts}
\usepackage{bm}

\def\beq{\begin{equation}}
\def\eeq{\end{equation}}
\def\bea{\begin{eqnarray}}
\def\eea{\end{eqnarray}}

\begin{document}

\title{Dynamics of Inflation and Dark Energy from $F(R, \mathcal{G})$ Gravity}
\author{S.D. Odintsov,$^{1,2,3}$\,\thanks{odintsov@ieec.uab.es}
V.K. Oikonomou,$^{4,5,6}$\,\thanks{v.k.oikonomou1979@gmail.com}
S. Banerjee,$^{7}$\thanks{sshreyabbanerjee@gmail.com },}
\affiliation{$^{1)}$ ICREA, Passeig Luis Companys, 23, 08010 Barcelona, Spain\\
$^{2)}$ Institute of Space Sciences (ICE,CSIC) C. Can Magrans s/n,
08193 Barcelona, Spain\\
$^{3)}$ Institute of Space Sciences of Catalonia (IEEC),
Barcelona, Spain\\
$^{4)}$ Department of Physics, Aristotle University of Thessaloniki, Thessaloniki 54124, Greece\\
$^{5)}$ Laboratory for Theoretical Cosmology, Tomsk State University
of Control Systems
and Radioelectronics (TUSUR), 634050 Tomsk, Russia\\
$^{6)}$ Tomsk State Pedagogical University, 634061 Tomsk, Russia\\
$^{7)}$ Tata Institute of Fundamental Research, India
}

\tolerance=5000

\begin{abstract}
In this work we study certain classes of $F(R,{\cal G})$ gravity which have appealing phenomenological features, with respect to the successful realization of the dark energy and of the inflationary era. Particularly, we discuss the general formalism and we demonstrate how several inflationary and dark energy evolutions can be described in the context of $F(R,{\cal G})$ gravity. Also we propose a unified model, in the context of which the early and late-time dynamics are controlled by the $F(R,{\cal G})$ gravity, thus producing inflation and the dark energy era, while the intermediate era is approximately identical with standard Einstein-Hilbert gravity. Also we calculate the power spectrum of the primordial curvature perturbations corresponding to the unified $F(R,{\cal G})$ gravity model we propose, which as we demonstrate is nearly scale invariant and compatible with the latest observational data constraints.
\end{abstract}


\maketitle

\section{Introduction}

The Universe at present time experiences an accelerating expansion, as standard observations based on Type IA supernovae indicate \cite{Riess:1998cb}. This accelerating expansion, dubbed dark energy era, is admittedly one of the most curious and surprising observations for our Universe, up to date. This is due to the fact that only a few people actually expected this late-time accelerating evolution, see for example \cite{weinberg}. A vast amount of works in theoretical cosmology is devoted towards describing the dark energy epoch, see for example \cite{Bamba:2012cp,Peebles:2002gy,Li:2011sd,Bamba:2010wb,Frieman:2008sn,Boehmer:2008av,Sahni:2006pa,Nojiri:2006gh,Elizalde:2004mq,Makarenko:2018blx}, and various descriptions in the literature exist on this issue, varying from quintessential or alternative approaches \cite{Capozziello:2003gx,Kamenshchik:2001cp,Carroll:1998zi,Capozziello:2002rd,Capozziello:2005ra}, to modified gravity descriptions see the reviews \cite{reviews1,reviews2,reviews3,reviews4,reviews5,reviews6}. Another quite intriguing epoch is the early-time era before the radiation dominated the expansion and evolution of the Universe. This is known as inflationary era, during which the Universe increased vastly its size in an exponential rate. The inflationary paradigm was introduced in the early 80's \cite{Guth:1980zm,Linde:1993cn,Linde:1983gd}, as a theoretical proposal that solved the shortcomings of the standard Big Bang cosmological model. The latest observations coming from Planck, severely constrained the spectral index of primordial curvature perturbations and the scalar-to-tensor ratio corresponding to this early epoch, however no direct observational proof that inflation occurred is given up to date. It may be possible that the bouncing cosmology paradigm \cite{bounce} may also be a viable description of the Universe at early times and even at late times.

In all the above cases, modified gravity \cite{reviews1,reviews2,reviews3,reviews4,reviews5,reviews6} in its various aspects, may successfully describe the early and late-time acceleration eras in a unified way, see for example \cite{Nojiri:2003ft} for a pioneer work towards this aspect. Modified gravity is based on generalizing Einstein's gravity at large scales and for various curvature limits. Actually it seems that at strong curvature limits and at large scales, Einstein's gravity fails to comply with the observational data. With regard to the dark energy epoch, the only way to describe an accelerating expansion in standard Einstein-Hilbert gravity, is to use a phantom fluid or field, and intriguingly enough, this phantom accelerating evolution leads to a crushing type singularity known as Big Rip \cite{Caldwell:2003vq}. In standard modified gravity descriptions, higher order curvature corrections are included in the standard Einstein-Hilbert gravitational action, and many examples of modified gravity exist in the literature, see the reviews \cite{reviews1,reviews2,reviews3,reviews4,reviews5,reviews6}. For the $f(R)$ gravity case, a well-known and quite successful model, from an observational point of view, is the  Starobinsky model first proposed in \cite{starobinsky} which is consistent with both local gravity constraints as well as large scale constraints. Also a quite appealing model that unifies late and early-time acceleration in the context of $f(R)$ gravity is the Nojiri-Odintsov model \cite{Nojiri:2003ft}. Alternative proposals to the $f(R)$ gravity theory use Lovelock invariants, such as the Gauss-Bonnet scalar $\mathcal{G}$, and several works exist that can successfully describe theoretically the dark energy epoch and also the inflationary epoch, see references \cite{Li:2007jm,Nojiri:2005jg,Nojiri:2005am,Cognola:2006eg,Elizalde:2010jx,Izumi:2014loa,Oikonomou:2016rrv,Kleidis:2017ftt,Oikonomou:2015qha,Escofet:2015gpa,Makarenko:2017vuk,new2,Makarenko:2016jsy} for an important stream of papers and recent works on the subject. Generalizations of the $f(R)$ and $f(\mathcal{G})$ gravity, are offered by higher order gravities \cite{Clifton:2006kc,Bogdanos:2009tn,Capozziello:2004us,Barrow:1988xh}, which use combinations of higher order curvature invariants constructed from the Ricci  and Riemann tensors, that is $R_{\mu\nu}R^{\mu\nu}$ and $R_{\mu\nu\sigma\rho}R^{\mu\nu\sigma\rho}$ respectively. Also theories which combine the Ricci and Gauss-Bonnet scalars also exist in the literature, namely, $F(R,{\cal G})$ gravity theories, and studies on this type of theories can be found in \cite{Elizalde:2010jx,Bamba:2009uf,DeLaurentis:2015fea,Benetti:2018zhv,DeFelice:2010sh,delaCruzDombriz:2011wn}. In this paper, we aim to realize several inflationary evolutions of main interest in the context of some classes of $F(R,{\cal G})$ gravity, and also we shall investigate how a late-time acceleration era can be realized from these theories. Our main interest will be on inflationary evolutions of physical interest, such as the quasi-de Sitter evolution, or variant forms of this. Finally, we discuss how the primordial curvature perturbations evolve in the context of $F(R,\mathcal{G})$ gravity, and we propose a phenomenologically interesting model, for which we calculate the power spectrum of the primordial curvature perturbations, and the corresponding spectral index. As we demonstrate, the viability of the model is achieved if the free variables are suitably constrained. Also, with regard to the dark energy epoch, we propose several phenomenologically viable models that may describe successfully this epoch.

The paper is organized as follows: In section 2 we present the fundamental features of $F(R,\mathcal{G})$ gravity in general, and we investigate how several inflationary scenarios can be generated by an appealing class of $F(R,\mathcal{G})$ gravity models. Also we investigate how a quasi-de Sitter inflationary evolution can be realized by a more general $F(R,\mathcal{G})$ gravity. In section 3, we present the realization of several dark energy scenarios in the context of a class of $F(R,\mathcal{G})$ gravity models, and in section 4 we propose a unification model that can both describe the inflationary and the dark energy epochs, and also during the intermediate epochs it can reduce to an Einstein-Hilbert model. Finally, the conclusions follow at the end of the article.

\section{$F(R,\mathcal{G})$ Gravity and Cosmological Solutions}

In this section we present the general formalism of $F(R,\mathcal{G})$, and we will demonstrate how the gravitational equations of motion can be used as a general reconstruction technique in order to realize several inflationary evolutions. We shall be interested in a certain class of $F(R,\mathcal{G})$ gravity, and we also present how a quasi-de Sitter evolution can be realized by a more general $F(R,\mathcal{G})$ gravity. The gravitational action of vacuum $F(R,\mathcal{G})$ gravity is equal to,
\begin{equation}
{\cal S}=\frac{1}{2\kappa}\int d^4x \sqrt{-g}F(R,{\cal G})\,,
   \label{action}
\end{equation}
where $\mathcal{G}$  stands for the Gauss-Bonnet invariant, defined as follows,
\begin{equation}
{\cal G}\equiv
R^2-4R_{\alpha\beta}R^{\alpha\beta}+R_{\alpha\beta\rho\sigma}
R^{\alpha\beta\rho\sigma}\, .
   \label{GBinvariant}
\end{equation}
By varying the gravitational action with respect to the metric tensor $g_{\mu \nu}$, we obtain the gravitational equations of motion, which are,
\begin{eqnarray}\label{eom}
&&G_{\mu \nu}= \frac{1}{ F_R} \Biggr[\nabla_\mu \nabla_\nu F_{R}-g_{\mu \nu} \Box F_{R}+2R \nabla_\mu \nabla_\nu F_{\cal G}\nonumber\\&&
-2g_{\mu \nu} R \Box F_{\cal G}-4R_\mu^{~\lambda} \nabla_\lambda \nabla_\nu F_{\cal G}-4R_\nu^{~\lambda} \nabla_\lambda \nabla_\mu F_{\cal G}
\nonumber \\&&
+4R_{\mu \nu} \Box F_{\cal G}+4 g_{\mu \nu} R^{\alpha \beta} \nabla_\alpha \nabla_\beta F_{\cal G}
+4R_{\mu \alpha \beta \nu} \nabla^\alpha \nabla^\beta F_{\cal G}
 \nonumber \\&&
-\frac{1}{2}\,g_{\mu \nu}\bigr( R  F_R + {\cal G}  F_{\cal G}- F(R,{\cal G})\bigr) \Biggr]\,,\\
&& 3\Box F_R + R F_R-2 F(R, {\cal G}) +R \left[ \Box F_{{\cal G}}+ 2\frac{{\cal G}}{R} F_{{\cal G}}\right]=0
\end{eqnarray}
In the following we shall assume that the geometric background is a flat Friedmann-Robertson-Walker (FRW) metric, with line element,
\begin{equation}
ds^{2}=-dt^{2}+a^{2}(t)(d{x}^{2}+d{y}^{2}+d{z}^{2})\, ,
\label{metric}
\end{equation}
so for the FRW metric (\ref{metric}), the gravitational equations (\ref{eom}) become,
\begin{eqnarray}\label{FRWH}
&& 2\dot{H}F_R + 8H \dot{H} \dot{F}_{\cal G}=H \dot{F}_R-\ddot{F}_R+4H^3\dot{F}_{\cal G}-4H^2\ddot{F}_{\cal G},\nonumber\\
\\
\label{energy}
&& 6H^2 F_R+24H^3\dot{F}_{\cal G}=F_RR-F(R,{\cal G})-6H\dot{F}_R+{\cal G}F_{\cal G} \,,\nonumber\\
\end{eqnarray}
where $R$ stands for the Ricci scalar, and $\mathcal{G}$ is the Gauss-Bonnet invariant, which for the FRW metric (\ref{metric}) these become equal to,
\begin{eqnarray}
R = 6 \left(2H^{2}+\dot H \right),\\
{\cal G} = 24H^{2} \left( H^{2}+\dot H \right)\, ,
\label{eq:R}\end{eqnarray}
and also $F_R=\frac{\partial F}{\partial R}$ and $F_{\mathcal{G}}=\frac{\partial F}{\partial \mathcal{G}}$. An appealing class of $F(R,{\cal G})$ gravity, with phenomenological interest, as we demonstrate at a later section, is the following \cite{Nojiri:2005jg},
 \begin{equation}
 F(R,{\cal G})=R+A {\cal{G}}^{\alpha}
 \label{form}
 \end{equation}
 where $\alpha$ is an even positive number and $A$ is an arbitrary constant of the theory. We shall assume that the parameter $A$ takes such values so that the term $A {\cal{G}}^{\alpha}$ is subdominant during the radiation domination epoch. Therefore we can assume that $A=(1/{\cal G}_i)^{\alpha}$, where ${\cal G}_i$ corresponds to the value of the Gauss-Bonnet curvature during the inflationary era. It is conceivable that the dominant term during inflation is also the Gauss-Bonnet curvature term $A {\cal{G}}^{\alpha}$, and also if $\alpha >1/2$, we have approximately the relation ${\cal G}\approx R^2$, so effectively a rational power of the Ricci scalar controls the evolution, which is valuable phenomenologically. At this point it is worth discussing the question how many degrees of freedom has the model (\ref{form}). This question was addressed in Ref. \cite{Bogdanos:2009tn}, and we repeat the argument of that work, so for a general theory with action,
\begin{equation}\label{actiongeneral}
\mathcal{S}=\int d^4x\sqrt{-g}f(R,P,Q)\, ,
\end{equation}
where $P=R_{\mu \nu}R^{\mu \nu}$ and $Q=R_{\mu \nu \sigma \beta }R^{\mu \nu \sigma \beta}$, the equation for the metric perturbations (gravitational wave modes) is,
\begin{equation}\label{gravitywaveseqn1}
 \Box h_f=m_s^2h_f\, ,
\end{equation}
\begin{equation}\label{gravitywaveseqn2}
\left(k^2+\frac{k^4}{m_{spin2}^2}\right)\bar{h}_{\mu \nu}=0\, ,
\end{equation}
where,
\begin{equation}\label{masses}
m_s^2=\frac{F_0}{F_{R0}+\frac{2}{3}(f_{P0}+f_{Q0})},\,\,\, m_{spin2}^2=-\frac{F_0}{f_{P0}+4f_{Q0}}\, ,
\end{equation}
and also $F_0=\frac{\partial f}{\partial R}\Big{|}_{R0}$, $F_{R0}=\frac{\partial F}{\partial R}\Big{|}_{R0}$, $f_{P0}=\frac{\partial f}{\partial P}\Big{|}_{R0}$ and finally $f_{Q0}=\frac{\partial f}{\partial Q}\Big{|}_{R0}$. For a general theory of the form (\ref{actiongeneral}), there are one massive scalar mode with mass $m_s^2$ corresponding to equation (\ref{gravitywaveseqn1}), one massless spin-2 mode corresponding to the term $k^2$ in Eq. (\ref{gravitywaveseqn2}) and finally one massive spin-2 ghost mode corresponding to the $k^4$ term in Eq. (\ref{gravitywaveseqn2}). Hence the general theory (\ref{actiongeneral}) has two more degree of freedom compared to the Einstein-Hilbert case. However, the model (\ref{form}) has only the massless spin-2 mode corresponding to the term $k^2$ in Eq. (\ref{gravitywaveseqn2}) and the massive scalar mode  with mass $m_s^2$, due to the fact that in this case $f_{Q0}=1$ and $f_{P0}=-4$. Hence the model (\ref{form}) has an additional degree of freedom in comparison to the Einstein-Hilbert gravity.

With regard to the inflationary era, we shall work assuming that the slow-roll conditions hold true, which in terms of the Hubble rate are expressed as follows,
 \begin{equation}
\dot{H}\ll H^2,,  \hspace{1cm} \ddot{H}\ll H \dot{H}\, .
\label{conditions}
\end{equation}
Also, for later use, the first two slow-roll indices of the Hubble slow-roll expansion are defined as follows,
\begin{equation}
\epsilon=-\frac{\dot{H}}{H^2}, \;\;\;\;\;\;\; \eta=-\frac{\ddot{H}}{2\;H\;\dot{H}}\, ,
\label{slowroll}
\end{equation}
and according to the slow-roll conditions, these two must be small during the inflationary era. Substituting Eq. \eqref{form} into the second Friedmann equation, and also by using Eq. \eqref{eq:R}, we get,
\begin{eqnarray}
-24^{\alpha}A(\alpha -1)H^{4\alpha}\left(1+\frac{\dot{H}}{H^2}\right)^{\alpha}+\frac{1}{(H^2+\dot{H})^2}\left(24^{\alpha}A(\alpha -1)\alpha H^{4\alpha}\left(1+\frac{\dot{H}}{H^2}\right)^{\alpha}(2\dot{H}(2H^2+\dot{H})+H\ddot{H})\right)=0
\end{eqnarray}
The above form is highly complicated, hence in order to get an analytical solution, we apply the slow-roll condition $\frac{\dot{H}}{H^2}\ll 1$ during the inflationary era. As a result, we can expand the terms like $(1+\frac{\dot{H}}{H^2})^{\alpha}$ up to first order. Simplifying the result, we finally get,
\begin{eqnarray}
24^{\alpha}A(\alpha-1)H^{4\alpha}\left(1+\alpha \frac{
\dot{H}}{H^2}\right)\left(-1+\alpha \frac{4\dot{H}H^2+2\dot{H}^2+H\ddot{H})}{H^4+2H^2\dot{H}}\right)=0\, .
\end{eqnarray}

The above equation can have three possible solutions, which are,
\begin{eqnarray}
(\alpha-1)&=&0 \\ {\rm or/and,}\ H^{4\alpha}(1+\alpha\frac{\dot{H}}{H^2})&=&0 \\ {\rm or/and,}\ H^4-(4\alpha -2)H^2\dot{H}-2\alpha\dot{H}^2-\alpha H\ddot{H}&=&0\, ,
\end{eqnarray}
and since, $H\neq0$, we get
 \begin{equation}
1+\alpha\frac{\dot{H}}{H^2}=0 \, ,
\label{sol2asd}
 \end{equation}
 or equivalently,
 \begin{equation}\label{sol2}
H(t)=-\frac{\alpha}{-t+c\alpha}\, ,
\end{equation}
 where $c$ is an arbitrary integration constant. The corresponding solution for the scale factor $a$ is given by,
 \begin{equation}
 a(t)=A_1(t-c\alpha)^{\alpha}\, ,
 \label{scale}
 \end{equation}
 where $A_1$ is an arbitrary integration constant. Finally the third solution can be found by solving,
 \begin{eqnarray}
 H^4-(4\alpha -2)H^2\dot{H}-2\alpha\dot{H}^2-\alpha H\ddot{H}=0\, .
 \end{eqnarray}
In order to obtain the evolution, we apply the slow-roll conditions, and the solution for $H$ is,
\begin{equation}
H(t)=\frac{2(1-2\alpha)}{t-2b(1-2\alpha)}\, ,
\label{sol3}
\end{equation}
where $b$ is again an integration constant. Accordingly, the scale factor is,
 \begin{equation}
 a(t)= A'(t -2 b (1 -2\alpha))^{(2 - 4\alpha)}\, ,
 \end{equation}
where $A'$ is an integration constant.

\subsection{Reconstructing the Universe Evolution: The $e$-foldings Number Approach}

As we have seen previously, the equations of motion are highly complicated, hence only approximate solutions can be obtained analytically. Therefore, in order to obtain exact solutions, without neglecting the Einstein term $R$, we shall employ the reconstruction technique firstly developed in \cite{Nojiri:2005pu,Bamba:2014daa} and then to $F(R)$ gravity in \cite{Bamba:2014wda,Rinaldi:2014gua}. Following the analysis of \cite{Bamba:2014wda}, we express all the variables as functions of the $e$-foldings number $N=\ln(a/a_0)$ (with $a_0$ being the scale factor at present time, which can be set equal to one for simplicity) instead of the cosmic time. Using the relation $dN=Hdt$ the second Friedmann equation appearing in  Eq. \eqref{FRWH} can be rewritten as follows,
\begin{equation}
H^2(N)(6F_R+24H^2(N)F_{\cal{GG}}{\cal G}'(N))=F_R R-F(R,{\cal G})-6H^2(N)F_{RR}R'(N)+{\cal G}(N)F_{\cal G}\, ,
\end{equation}
where we have used Eq. \eqref{eq:R}, expressed as follows in terms of the $e$-foldings number,
\begin{eqnarray}
R(N) = 6 \left(2H^{2}(N)+H(N)H'(N) \right),\\
{\cal G}(N) = 24H^{2}(N) \left( H^{2}(N)+H(N)H'(N) \right)\, .
\label{eq:RR}\end{eqnarray}
In the above equations, the prime indicates differentiation with respect to the $e$-foldings number. Eq. (\ref{eq:RR}) can be rewritten as follows,
\begin{equation}
24H^4(N)F_{\cal{GG}}{\cal G}'(N)+6H^2(N)(F_R+F_{RR}R'(N))+F(R,{\cal G})-F_RR(N)-{\cal G}(N)F_{\cal G}=0\, .
\label{soln}
\end{equation}
Solving with respect to $H^2(N)$ we get,
\begin{equation}
H^2(N)=\frac{-6(F_R+F_{RR}R'(N))+\sqrt{(6(F_R+F_{RR}R'(N)))^2-96F_{\cal{GG}}{\cal G}'(N)(F(R,{\cal G})-F_RR(N)-{\cal G}(N)F_{\cal G})}}{48F_{\cal{GG}}{\cal G}'(N)}\, .
\label{Hubble}
\end{equation}
Thus we see that $H$ is now a function of $R,\ {\cal G}, F(R,{\cal G})$ and of higher derivatives. Using the expressions for slow-roll parameters defined previously, in terms of $H$ as given in Eq. \eqref{slowroll}, one can now express all the slow-roll indices as functions of $R,\ {\cal G}, F(R,{\cal G})$ and of higher derivatives (all being functions of the $e$-foldings number $N$).

Using the relation $dN=Hdt$, we can rewrite the slow roll parameters such that $t$ is replaced by $N$ in the following way,
\begin{equation}
\epsilon=-\frac{H(N)H'(N)}{H^2(N)};\ \ \eta=-\frac{H''(N)H(N)+H'^2(N)}{H(N)H'(N)}
\label{srH}
\end{equation}
Also from Eq. \eqref{eq:RR}, we have,
\begin{eqnarray}
H(N)H'(N)&=&\frac{\cal{G}(N)}{24 H^2(N)}-H^2(N) \\
H'^2(N)+H(N)H''(N)&=& \frac{\cal{G}'(N)}{24H^2(N)}-4\left(\frac{\cal{G}(N)}{24H^2(N)}-H^2(N)\right)-\frac{2}{H^2(N)}\left(\frac{\cal{G}(N)}{24H^2(N)}-H^2(N)\right)^2
\end{eqnarray}
Using this information and substituting the expression for $H^2(N)$ from Eq. \eqref{Hubble}, we may express the slow-roll parameters as follows,
\begin{eqnarray}
\epsilon &=& -\frac{\cal{G}(N)}{24\left(\frac{-6(F_R+F_{RR}R'(N))+\sqrt{(6(F_R+F_{RR}R'(N)))^2-96F_{\cal{GG}}{\cal G}'(N)(F(R,{\cal G})-F_RR(N)-{\cal G}(N)F_{\cal G})}}{48F_{\cal{GG}}{\cal G}'(N)}\right)^2}+1 \\
\eta &=& -\frac{\cal{G}'(N)}{{\cal{G}}(N)-24(\frac{-6(F_R+F_{RR}R'(N))+\sqrt{(6(F_R+F_{RR}R'(N)))^2-96F_{\cal{GG}}{\cal G}'(N)(F(R,{\cal G})-F_RR(N)-{\cal G}(N)F_{\cal G})}}{48F_{\cal{GG}}{\cal G}'(N)})^2} \nonumber \\ && ~~~ +2+2\frac{\cal{G}(N)}{24(\frac{-6(F_R+F_{RR}R'(N))+\sqrt{(6(F_R+F_{RR}R'(N)))^2-96F_{\cal{GG}}{\cal G}'(N)(F(R,{\cal G})-F_RR(N)-{\cal G}(N)F_{\cal G})}}{48F_{\cal{GG}}{\cal G}'(N)})^2}
\end{eqnarray}
Thus we can see that given the $F(R,{\cal G})$ function and also the functional form of $H^2(N)$, one can find the expressions for the slow-roll parameters. If assumed that the $F(R,\mathcal{G})$ gravity acts as a perfect fluid, one may calculate the spectral index of the primordial curvature perturbations and of the scalar-to-tensor ratio by using the usual expressions corresponding to a canonical scalar field. However, we need to stress that a more formal approach requires the calculation of the spectral index directly by calculating the power spectrum of primordial curvature perturbations, as we formally do in a later section for a phenomenologically appealing $F(R,\mathcal{G})$ gravity.

What we have at hand up to now in this section, is a reconstruction method that can enable us, given the Hubble rate as a function of the $e$-foldings number, to find the $F(R,\mathcal{G})$ function that may realize such an evolution, as a function of the $e$-foldings of course. The reconstruction method consists of the following steps, firstly we assume a specifically chosen functional form of the Hubble rate, for example,
\begin{equation}
H^2(N)=P(N)\, .
\end{equation}
Then by using Eq. \eqref{eq:RR}, we obtain,
\begin{eqnarray}
R=12P(N)+3P'(N);\quad {\cal G}=24P^2(N)+12P(N)P'(N)\, .
\label{GP}
\end{eqnarray}
The Friedmann equation can now be written as follows,
\begin{eqnarray}
&&24P^2(N)(48P(N)P'(N)+12P(N)P''(N)+12P'^2(N))F_{\cal{GG}}+6P(N)(12P'(N)+3P''(N))F_{RR}\nonumber \\ &&~~~~~~~~~~~~~~~~~~~-(24 P^2(N) +12P(N)P'(N))F_{\cal G}-(6P(N)+P'(N))F_R+F(R,{\cal G})=0
\label{soln1}
\end{eqnarray}
Thus, by appropriately choosing the function $P(N)$ and then by using the functional relation between $R,\ {\cal G}$ and the number of e-foldings, the resulting equation becomes a second order differential equation for the function $F(R,{\cal G})$, whose solution can be obtained in principle. On the other hand, we can also assume a particular form of $F(R,{\cal G})$ and solve the above equation for $P(N)$, however in general this might be technically more difficult.

In order to demonstrate how the method works in practise, let us present some illustrative and relatively simple examples. We start off by assuming that the Hubble rate has the following form \cite{Bamba:2014wda},
\begin{equation}
H^2(N)=P_0 N=P(N)\, ,
\label{P}
\end{equation}
which in terms of the cosmic time is an exact de Sitter evolution $H(t)\sim e^{P_0 t}$. We shall seek which $F(R,\mathcal{G})$ gravity of the form,
\begin{equation}
F(R,{\cal G})=R+F({\cal G})
\label{F(R,G)}
\end{equation}
 can realize the above evolution. Using Eqs. \eqref{P} and \eqref{GP}, we obtain,
\begin{eqnarray}
N({\cal G})=\frac{-1}{4}+\frac{1}{4}\left(1-\frac{8{\cal G}}{P_0^2}\right)^{1/2} \\
P({\cal G})=\frac{P_0}{4}\left(-1+\left(1-\frac{8{\cal G}}{P_0^2}\right)^{1/2}\right)
\end{eqnarray}
Substituting the above forms into Eqs. \eqref{soln1}, we find that the resulting form of the differential equation is still too complicated. In order to simplify the resulting equations, we shall assume that the slow-roll condition holds true, thus by neglecting $H'$ terms in comparison to $H^2$ terms, we obtain,
\begin{equation}
{\cal G}=24 P^2(N)=24P_0^2N^2,\quad R=12P(N)\, ,
\label{g}
\end{equation}
Using the above form for the Gauss-Bonnet scalar, we get,
\begin{eqnarray}
N({\cal G})&=&\frac{\cal G}{24^{1/2}P_0} \\
P({\cal G})=\frac{{\cal G}^{1/2}}{24^{1/2}}
\end{eqnarray}
Thus Eq. \eqref{soln1} is simplified as follows,
\begin{eqnarray}
24P^2(N)\times 48P(N)P'(N)F_{\cal{GG}}+6P(N)\times 12P'(N)F_{RR}-24 P^2(N)F_{\cal G}-6P(N)F_R+F(R,{\cal G})=0
\end{eqnarray}
Accordingly, by using Eq. \eqref{F(R,G)}, we get,
\begin{equation}
24^{1/2}P_0 {\cal G}^{3/2}F_{\cal{GG}}-{\cal G}F_{\cal G}+6 {\cal G}^{1/2}/24^{1/2}+F({\cal G})=0
\end{equation}
The solution of the above equation is the following,
\begin{eqnarray}\label{solmeijer}
F({\cal G})&=&F_0\frac{1}{6}{\cal G}+F_1G_{1\,1}^{0\,0}(-\sqrt{\frac{{\cal G}}{6}} \Big{|}0,2) \nonumber \\ &&~~~-\frac{1}{6}e^{-\sqrt{\frac{{\cal G}}{6}}}\left(36G_{1\,1}^{0\,0}(-\sqrt{\frac{{\cal G}}{6}} \Big{|}0,2)+6\sqrt{6}\sqrt{\cal G}G_{1\,1}^{0\,0}(-\sqrt{\frac{{\cal G}}{6}} \Big{|}0,2)\right)\, .
\end{eqnarray}
Here $F_0$ and $F_1$ are integration constants chosen such that the Gauss-Bonnet term has minimum effect as the Universe enters the radiation domination era, and $G_{m\,n}^{p\,q}(z)$ is the Meijer $G$ function. It should be noted that the first term in Eq. (\ref{solmeijer}) does not affect the dynamics of the cosmological system, since it contributes a total divergence. Actually, only the second and third terms affect the dynamics. A detailed analysis of the Meijer $G$ function in the large ${\cal G}$ regime, indicates that the second term in the solution (\ref{solmeijer}) is dominant, thus we have during the inflationary era,
\begin{eqnarray}
F({\cal G})&=&F_1 G_{1\,1}^{0\,0}(-\sqrt{\frac{{\cal G}}{6}} \Big{|}0,2)\, .
\label{f(g)}
\end{eqnarray}
This practically is the result of the reconstruction method we presented, so the exact de Sitter evolution of Eq. (\ref{P}) is realized during the slow-roll inflationary era by the $F(R,\mathcal{G})$ function $F(R,\mathcal{G})=R+F_1G_{1\,1}^{0\,0}(-\sqrt{\frac{{\cal G}}{6}} \Big{|}0,2)$.


Let us consider another illustrative and relatively simple example, in which case the Hubble rate as a function of the $e$-foldings number $N$ is,
\begin{equation}
H^2(N)=P(N)=P_2e^{\beta N}
\label{HH}
\end{equation}
with $P_2>0$. Such exponential behavior gives rise to a power law scale factor as a function of the cosmic time, which can give rise to an inflationary evolution. In this case we have,
\begin{eqnarray}
N({\cal G}) &=&\frac{1}{2\beta}\ln \frac{{\cal G}}{12 P_2^2(2+\beta)} \\
P({\cal G}) &=& \frac{{\cal G}^{1/2}}{(12 P_2^2(2+\beta))^{1/2}}\, ,
\end{eqnarray}
and in effect, the differential equation for the $F(R,\mathcal{G})$  is,
\begin{equation}
\frac{2\beta{\cal G}^2}{(2+\beta)^2}(3+2\beta)F_{\cal{GG}}+3^{1/2}\frac{{\cal G}^{1/2}}{(2+\beta)^{1/2}}-{\cal G}F_{\cal G}+F({\cal G})=0\, ,
\end{equation}
which can be analytically solved and the solution is,
\begin{equation}
F({\cal G})=\frac{4 b}{a-2}{\cal G}^{1/2}+F_0{\cal G}^{1/a}+F_1{\cal G}\, ,
\end{equation}
where,
\begin{equation}
a=\frac{2\beta(3+2\beta)}{(2+\beta)^2},\quad \quad b=\frac{3^{1/2}}{(2+\beta)^{1/2}}\, .
\end{equation}
Here $F_0,\ F_1$ are integration constants. As it can be seen from the functional form of the solution, the first term is $\sim {\cal G}^{1/2}$, hence it behaves like general relativity. The dominant term in the large Gauss-Bonnet curvature regime is the second term, and for it to be the dominating one, we should have $a<1$. For example, if $a=0.5$, then we get the simplest modification with ${\cal G}^2$. In this case, $\beta=-1.5\ {\rm or}\ 0.38$.


The reconstruction method we presented works if the Hubble rate $H(N)$ is given, but the method can work the other way round, that is to fix the $F(R,\mathcal{G})$ gravity and from it to find the Hubble rate $H(N)$. Indeed, by choosing a particular form of $F({\cal G})$ as was done earlier, substituting it into Eq. \eqref{soln1} and using Eq. \eqref{GP} we get a differential equation for $P(N)$, solving which yields the form of $H^2(N)$. Let us demonstrate this by choosing the functional form of $F(R,{\cal G})$ to be,
\begin{equation}
F(R,{\cal G})=R+A{\cal G}^{\alpha}
\label{f(R,G)}
\end{equation}
Using the above form, we get the following second order differential equation for the function $P(N)$,
\begin{eqnarray}
&& 24^{\alpha+1}AP^{2\alpha+2}(N)\alpha^2(48P(N)P'(N)+12P(N)P''(N)+12P'^2)(1+\frac{\alpha P'(N)}{P(N)})\nonumber \\ &&~~~~~~~-24^{\alpha}AP^{2\alpha}(N)(24 P^2(N)+12P(N)P'(N))(1+\frac{\alpha P'(N)}{P(N)})+24^{\alpha}AP^{2\alpha}(N)(1+\frac{\alpha P'(N)}{P(N)})+6P(N)=0\, ,
\end{eqnarray}
where we used the approximation $(1+P'(N)/P(N))^{\alpha}\approx(1+\alpha P'(N)/P(N))$, which are imposed by the slow-roll approximation. Solving the above differential equation with respect to $P(N)$, we get the form of $H^2(N)$ which is,
\begin{equation}
P(N)=P_0e^{-2N/\beta}
\end{equation}
with $\beta=2\alpha$. Thus, constraining $\beta$, one can in turn constrain $\alpha$.

\subsection{Quasi-de Sitter Evolution in the Context of a General $F(R,\mathcal{G})$ Gravity}

In principle, in the context of $F(R,\mathcal{G})$ gravity, any cosmological evolution can be realized, however the resulting differential equations to solve are not always easy to tackle. Nevertheless, simple cosmological solutions can easily be handled, and a particularly interesting and phenomenologically valuable evolution, is the quasi-de Sitter evolution, in which case the Hubble rate is,
\begin{equation}\label{qsdesitter}
H(t)=H_0-H_it\, ,
\end{equation}
where $H_0$ and $H_i$ are dimensionful parameters of the theory. In general, the latter two parameters are controlled by the modified gravity governing the evolution, and are constrained by the observational data. For example, in the case of $R^2$ gravity, a concise treatment on the allowed values for $H_0$ and $H_i$ can be found in Ref. \cite{Odintsov:2015gba}, but for the purposes of this paper we shall assume that $H_0\gg H_i$, which is justified by the conditions of having a nearly de Sitter evolution. We shall be interested in finding which $F(R,\mathcal{G})$ gravity may generate the quasi de Sitter evolution (\ref{qsdesitter}), so we shall use the reconstruction method of Ref. \cite{Bamba:2009uf}. We rewrite the gravitational action as follows,
\begin{equation}\label{frgactionnew}
\mathcal{S}=\int d^4x\sqrt{-g}\left(P(t)R+Z(t)\mathcal{G}+Q(t) \right)\, ,
\end{equation}
and upon variation with respect to $t$, which is an auxiliary scalar field identified with the cosmic time, we get,
\begin{equation}\label{auxiliary1}
P'(t)R+Z'(t)\mathcal{G}+Q'(t)=0\, ,
\end{equation}
where the ``prime'' denotes differentiation with respect to the auxiliary scalar ``t''. From Eq. (\ref{auxiliary1}) one can determine the relation $t=t(R,\mathcal{G})$, then if someone finds the functions $P(t)$, $Q(t)$ and $Z(t)$, then the $F(R,\mathcal{G})$ function is found by substituting $t=t(R,\mathcal{G})$ found from Eq. (\ref{auxiliary1}). By using the equation of motion (\ref{FRWH}), we find that the differential equations which determine the functions $P(t)$, $Q(t)$ and $Z(t)$, are the following,
\begin{equation}\label{diffeqneakyro1}
P''(t)+4\dot{g}^2(t)Z''(t)-\dot{g}(t)P'(t)+(8\dot{g}\ddot{g}-4\dot{g}^3(t))Z'(t)+2\ddot{g}P(t)=0\, ,
\end{equation}
\begin{equation}\label{diffeqneakyro2}
Q(t)=-24\dot{g}^3(t)-6\dot{g}^2(t)P(t)-6\dot{g}P'(t)\, .
\end{equation}
Thus by determining the functions $P(t)$, $Q(t)$ and $Z(t)$, which can be found by directly solving the differential equation (\ref{diffeqneakyro1}), the explicit form of the $F(R,\mathcal{G})$ gravity can be found. Let us apply this method for the quasi-de Sitter evolution (\ref{qsdesitter}), so we assume for simplicity that the function $Z(t)$ has the following form,
\begin{equation}\label{zt}
Z(t)=\beta e^{b t}\, ,
\end{equation}
where $b$ and $\beta$ are arbitrary real parameters. Then, by substituting $Z(t)$ in Eq. (\ref{diffeqneakyro1}), we obtain the differential equation that will yield the function $P(t)$. However, the resulting differential equation is very complicated to even quote it here, and also to solve it, so some leading order approximation is needed. Emphasizing at early times, the differential equation (\ref{diffeqneakyro1}), becomes at leading order,
\begin{equation}\label{leadingordereqn}
P''(t)-H_0P'(t)-2H_iP(t)+\mathcal{S}(t)\simeq 0\, ,
\end{equation}
where the function $\mathcal{S}(t)$ is equal to,
\begin{equation}\label{functions}
\mathcal{S}(t)=4 b^2 \beta  H_0^2 e^{b t}-8 b \beta  H_0^4 e^{b t}+16 b \beta  H_0 H_i^2 e^{b t}\, .
\end{equation}
The differential equation (\ref{leadingordereqn}) is easy to solve, so the resulting solution is,
\begin{equation}\label{ptsolution}
P(t)=\mathcal{C}_1\frac{1}{2} (H_0-\mu ) e^{\frac{1}{2} t (H_0-\mu )}+\mathcal{C}_2\frac{1}{2} (H_0+\mu ) e^{\frac{1}{2} t (H_0+\mu )}-\frac{4 b^2 \beta  H_0 e^{b t} \left(b H_0-2 H_0^3+4 H_i^2\right)}{b^2-b H_0-2 H_i}\, ,
\end{equation}
where $\mathcal{C}_1$ and $\mathcal{C}_2$ are integration constants, and $\mu=\sqrt{H_0^2+8 H_i}$. Substituting Eqs. (\ref{zt}) and (\ref{ptsolution}) into Eq. (\ref{diffeqneakyro2}), we obtain the function $Q(t)$, which at leading order is,
\begin{align}\label{functionq}
& Q(t)\simeq \frac{24 b^2 \beta  H_0^2 e^{b t} \left(b H_0-2 H_0^3+4 H_i^2\right)}{b^2-b H_0-2 H_i}-24 b \beta  H_0^3 e^{b t}-6 H_0^2 \\ \notag &
-3 \mathcal{C}_1 H_0 (H_0-\mu ) e^{\frac{1}{2} t (H_0-\mu )}-3 \mathcal{C}_2 H_0 (H_0+\mu ) e^{\frac{1}{2} t (H_0+\mu )}\, .
\end{align}
Finally, by combining Eqs. (\ref{zt}), (\ref{ptsolution}), (\ref{diffeqneakyro2}) and (\ref{auxiliary1}), we can obtain the relation $t=t(R,\mathcal{G})$. By substituting the resulting expression for $t=t(R,\mathcal{G})$ in $F(R,\mathcal{G})=P(t)R+Z(t)\mathcal{G}+Q(t)$, we obtain the approximate resulting form of the $F(R,\mathcal{G})$ gravity which realizes the quasi-de Sitter evolution (\ref{qsdesitter}), which is,
\begin{equation}\label{frgfinal}
F(R,\mathcal{G})\simeq -\frac{-\Gamma_2+\Gamma_4+\Gamma_3 G-\Gamma_1 R}{-\mathcal{C}_2 R+\gamma -\mathcal{G}\beta  }-6 H_0^2\, ,
\end{equation}
where the parameters $\Gamma_i$, $i=1,...,4$ and $\gamma$ can be found in the Appendix. We need to note that there is in principle much freedom in choosing the function $Z(t)$ and determining the resulting $F(R,\mathcal{G})$ gravity, therefore it is possible that functionally different $F(R,\mathcal{G})$ gravities may realize the same cosmological evolution, at least in the vacuum case. However, the complicated equations require extensive simplifications, so from now on we shall consider $F(R,\mathcal{G})$ gravities of the form $F(R,\mathcal{G})=R+f(\mathcal{G})$.

\section{Dark energy from $F(R,{\cal G})$ Gravity}

Having discussed the realizations of inflationary evolutions in the context of $F(R,{\cal G})$ gravity, in this section we shall present several viable models of $F(R,{\cal G})$ which may successfully generate a dark energy era. Some simple but in general phenomenologically viable models are given below,
\begin{eqnarray}
{\rm I)}\ F(R,{\cal G})&=&\frac{R}{2}-F_0(1-e^{{\cal{G}}/{\cal{G}}_0}); \\
F_{\cal G}&=&\frac{F_0}{{\cal G}_0}e^{{\cal{G}}/{\cal{G}}_0};\quad F_{\cal{GG}}=\frac{F_0}{{\cal G}_0^2}e^{{\cal{G}}/{\cal{G}}_0} \\
{\rm II)}\ F(R,{\cal G})&=&\frac{R}{2}-F_0(1-e^{-|{\cal{G}}|/{\cal{G}}_0}); \\
F_{\cal G}&=&-\frac{F_0}{{\cal G}_0}e^{-|{\cal{G}}|/{\cal{G}}_0};\quad F_{\cal{GG}}=\frac{F_0}{{\cal G}_0^2}e^{-|{\cal{G}}|/{\cal{G}}_0} \\
\label{de}
\end{eqnarray}
where ${\cal G}_0$ corresponds to the present value of the Gauss-Bonnet scalar and $F_0$ is a constant which will be chosen to be equal to the present value of the cosmological constant. In this work we will mainly be interested in reproducing the dark energy era, using exponential $F({\cal G})$ gravities, since this functional form of $F(R,{\cal G})$ gravity is relatively simple and it can easily satisfy the solar system constraints.

Since ${\cal G}=24H^2(H^2+\dot{H})\equiv 24H^2\ddot{a}/a$,  the term $\ddot{a}$ changes sign, that is, goes from negative to positive, as the Universe transits from the decelerating to the accelerating phase, therefore ${\cal G}$ also goes from negative to positive. Thus in the model I in Eq. (\ref{de}), the change in sign of ${\cal G}$ affects the exponential part whereas in the model II, since we have $|{\cal G}|$, the change in the sign has no direct effect.

Neglecting radiation and matter fluids, thus considering the models in vacuum, de-Sitter points correspond to a constant  Hubble rate value, say $H=H_0$ (which will correspond to the present value of Hubble constant in case we are dealing with late-time acceleration), and the corresponding equation that determines the de Sitter points is given by,
\begin{eqnarray}
\label{deSi}
3H_0^2={\cal{G}}_0 F_{\cal G}({\cal G}_0)-F({\cal G}_0)\, ,
\end{eqnarray}
where ${\cal G}_0=24H_0^4$. Let us consider first the model I in Eq. (\ref{de}), so by using  Eq. \eqref{de}, we get
\begin{equation}
F_0=3H_0^2
\end{equation}
In order to study the stability of the de Sitter point, let us consider a linear perturbation around the de Sitter point $H=H_0+\delta H$, so we get the perturbation equation which is \cite{DeFelice:2010sh},
\begin{eqnarray}
\label{desitter}
\delta{\ddot{H}}_0+3H_0\delta{\dot{H}}_0 +\left[ \frac{1}{96H_0^6
F_{\cal{GG}}(H_0)}-4\right] H_0^2 \delta H_0=0\,.
\end{eqnarray}
which can be solved to yield the solution,
\begin{eqnarray}
\delta H_0=c_1 e^{\lambda_+ t}+c_2 e^{\lambda_- t}\,,
\qquad
\lambda_\pm=\frac{3H_0}{2} \left[ -1 \pm
\sqrt{1-\frac49 \left( \frac{1}{96H_0^6
F_{\cal{GG}}(H_0)}-4 \right)} \right]\, ,
\end{eqnarray}
where $c_1$ and $c_2$ are integration constants \cite{DeFelice:2010sh}.
Thus in order for the de Sitter points to be stable, the following conditions need to be satisfied,
\begin{eqnarray}
\label{decon}
0<H_0^6 F_{\cal{GG}} (H_0) <1/384\, .
\end{eqnarray}\
Hence the stability depends on the sign and the magnitude of $F_{\cal{GG}}$. For the model I $F({\cal G})$, one gets the form of $\lambda=3H_0/2(-1\pm \sqrt{2.4})$ and the quantity $H_0^6 F_{\cal{GG}} (H_0)=3e^{1}/(24^2)>1/384$. Thus, this model has slightly unstable de Sitter points. This implies that the late-time de Sitter attractor is unstable, however from a phenomenological point of view, this is not a serious issue, since we do not know how the Universe will evolve in the future, that is, if the late-time de Sitter attractor is stable or not.

For the model II, we have,
\begin{equation}
F_0=\frac{3H_0^2}{1-2e^{-1}}\, ,
\end{equation}
and in order to have stability, the following condition must hold true $F_{\cal{GG}}>0$, so in this case we must have $\lambda=3H_0/2(-1\pm \sqrt{2})$ and also $H_0^6 F_{\cal{GG}} (H_0)=3e^{-1}/((1-2e^{-1})\times 24^2)>1/384$. As a result, in this case too we have an unstable de Sitter point as a solution.

\subsection{Study of the Equation of State}

Let us investigate the behavior of the effective equation of state (EoS) for the above models, so we rewrite the equations of motion as follows,
\begin{eqnarray}
\label{GBfrw1}
\frac{3}{\kappa^2}H^2 &=& \rho_{\cal G} + \rho_M \ ,\\
\frac{1}{\kappa^2}\left(2\dot H + 3H^2\right) &=&  - p_{\cal G} - p_M \ .
\end{eqnarray}
where we have included the presence of matter, and $\rho_{\cal G}$ and $p_{\cal G}$ are given by,
\begin{eqnarray}
\label{GBfrw2}
\rho_{\cal G} &=& {\cal G}F_{\cal G} - F({\cal G}) - 24 \dot{\cal G} F_{\cal{GG}} H^3 \ ,\nonumber \\
p_{\cal G} &=& -  {\cal G}F_{\cal G} + F({\cal G}) + 24 \dot{\cal G}F_{\cal{GG}} H^3 + 8\dot{\cal G}^2F_{\cal{GGG}}H^2
\nonumber \\
&& - 192 F_{\cal{GG}}\left( - 8 H^3 \dot{H} \ddot{H} - 6 H^2 \dot{H}^3 - H^4
\dddot{H} \right. \left. - 3H^5 \ddot{H} - 18 H^4 \dot{H}^2 + 4
H^6 \dot{H} \right)\, .
\end{eqnarray}
Thus the EoS parameter corresponding to the geometric dark energy is $w=p_{\cal G}/\rho_{\cal G}$. At the de Sitter point, by substituting $H=H_0$ and ${\cal G}={\cal G}_0$ as constants, we get $\rho_{\cal G}=-p_{\cal G}$, as was expected. Let us consider model I, so in order to get the evolution for the Hubble parameter, and in effect to obtain the EoS corresponding to late times (so for ${\cal G}<{\cal G}_0$) we need to solve the Friedmann equation for the model I of Eq. \eqref{de}. Since the resulting equation is very complicated, we apply certain approximations by using the fact that we are interested for low curvature terms, so one can expand the exponential term and keep up to the first non-zero contributions of the Gauss-Bonnet term in the action we get $F(R,{\cal G})=\frac{R}{2}+F_0(\frac{\cal G}{{\cal{G}}_0}+\frac12(\frac{\cal G}{{\cal{G}}_0})^2)$. Then substituting the above form in the Friedmann equation and by neglecting the Einstein term one finds that the approximate solution for Hubble parameter is given by,
\begin{equation}
H=\frac{A}{t}\, ,
\end{equation}
which denotes an accelerating Universe with an approximate power law scale factor. The EoS for any modified gravity is equal to \cite{reviews1},
 \begin{equation}
 w=-1-\frac{2}{3}\frac{\dot{H}}{H^2}
 \end{equation}
which in the case at hand becomes equal to $w=-1+\frac{2}{3A}$. Therefore, depending on the sign of the parameter $A$, one may have quintessential acceleration (for $A>0$ and for values which render $w<-1/3$) or a phantom evolution (for $A<0$). Also, for large values of $A$, the evolution is approximately a de Sitter evolution. The same results are obtained for the model II in Eq. (\ref{de}), so we omit the study of the model II for brevity.

\subsection{Stability of Matter-Radiation Points}

Let us now focus on the stability of matter and radiation points solutions. In order to study the stability, we consider the following equation  \cite{DeFelice:2010sh},
\begin{equation}
\label{background}
3H^2={\cal G} F_{\cal G}-F({\cal G})-24H^3 F_{\cal{GG}} \dot{\cal G}+\rho_M\,.
\end{equation}
We shall study perturbations of linear order and of the following form,
\begin{equation}
H=H^{(b)} (1+\delta_H)\,,\quad
\rho_M=\rho_M^{(b)} (1+\delta_M)\,.
\end{equation}
By keeping linear order perturbations, the equation that determines the stability of matter and radiation points is given below \cite{DeFelice:2010sh},
\begin{eqnarray}
\label{geper}
\ddot{\delta_H}&&+\left[ 3-\frac{6}{p}+\frac{96H^4 F_{\cal{GGG}}}
{f_{\cal{GG}}}
\frac{1-p}{p^2} \right] H \dot{\delta_H}
+\left[ \frac{21(1-p)}{p^2}+\frac{1}{96H^6 f_{\cal{GG}}}-4+
\frac{96H^4 f_{\cal{GGG}}}{f_{\cal{GG}}} \left( 4-\frac{3}{p} \right)
\frac{1-p}{p^2} \right] H^2 \delta_H \nonumber \\
&&=\frac{\delta_M}{192 H^6 f_{\cal{GG}}}H^2
\end{eqnarray}
where we have assumed $a\propto t^{p}$. Then the stability conditions are the following \cite{DeFelice:2010sh},
\begin{eqnarray}
F_{\cal{GG}}>0 \ {\rm for}\ {\cal G} \leq {\cal G}_0 ; \ F_{\cal{GG}}\longrightarrow 0+ \ {\rm for} \ |{\cal G}|\longrightarrow \infty \, .
\end{eqnarray}
In addition, the modified gravity term originating from the functional form of the $F(R,\mathcal{G})$ gravity, should also vanish or become subdominant during the matter and radiation domination epochs, so $F(R,{\cal G}) \longrightarrow R/2$ as $|{\cal G}|\gg {\cal G}_0$. Let us firstly consider the model I in Eq. (\ref{de}), so we need to stress that ${\cal G}$ is negative during the matter and radiation dominated epochs, hence the signs of ${\cal G}$ and ${\cal G}_0$ are opposite in the exponent of the model I in Eq. \eqref{de}. In effect, an exponentially decreasing function is obtained in the regime $|{\cal G}|>> {\cal G}_0$. As can be seen from Eq. \eqref{de}, and for the model I, for ${\cal G}>> {\cal G}_0$, $F_{\cal{GG}}\rightarrow 0+$. Also in the large curvature regime, $F(R,{\cal G})=R/2-F_0\approx R/2 $, thus we obtain a behavior similar to that of Einstein-Hilbert gravity. Let us now consider model II in Eq. (\ref{de}), and since the absolute value of ${\cal G}$ appears in the exponent, we have  a decreasing function of ${\cal G}$, irrespective of the sign of ${\cal G}$, thus satisfying all the above conditions. Therefore, as a result, both the models appearing in Eq. (\ref{de}) can describe an accelerating late time evolution, but also can ensure an Einstein-Hilbert-like evolution during the radiation and matter domination epochs.

\subsection{Reconstruction of Several Cosmic Scenarios}

Before closing this section, let us demonstrate how to realize several cosmological scenarios by using the models (\ref{de}). Firstly, for a de Sitter Universe, the Gauss-Bonnet and the Ricci scalars are constant (and can be set equal to their present value), so we have,
\begin{equation}
R=R_0 \quad \quad {\cal G}={\cal G}_0=\frac16 R_0^2
\end{equation}
Following the analysis of Ref. \cite{Cognola:2006eg}, the resulting gravitational equations of motion take the following form (after taking the trace and writing $F(R,{\cal G})=R+F({\cal G})$),
\begin{equation}
{\cal G}_0F_{\cal G}({\cal G}_0)-F({\cal G}_0)=\frac12 R_0\, .
\end{equation}
In the $\Lambda$CDM model, we have $R_0=4\Lambda$, thus the effective cosmological constant in this theory is equal to,
\begin{equation}
\Lambda_{eff}=\frac12({\cal G}_0F_{\cal G}({\cal G}_0)-F({\cal G}_0))\, .
\end{equation}
Hence, for the models I we have,
\begin{equation}
\Lambda_{eff}=\frac12 F_0\, ,
\end{equation}
while for the model II, we have,
\begin{equation}
\Lambda_{eff}=\frac12F_0(1-2e^{-1})\, .
\end{equation}
In order to obtain several cosmic evolutions realizations, we shall apply the reconstruction techniques we presented in earlier sections. In this way, we shall be able to realize several cosmic scenarios so by following the analysis of \cite{Nojiri:2009kx}, the Friedmann equation is,
\begin{eqnarray}
6 P(N({\cal G}))&&+24(48(P(N({\cal G})))^3P'(N({\cal G}))+12(P(N({\cal G})))^2(P'(N({\cal G})))^2+12 (P(N({\cal G})))^3P''(N({\cal G})))F_{\cal{GG}} \nonumber \\ &&~~~~~~~ -(24(P(N({\cal G})))^2+12P(N({\cal G}))P'(N({\cal G})))F_{\cal G}+F({\cal G})-\sum_{i}\rho_{0i}a_0^{-3(1+w_i)}e^{-3(1+w_i)N({\cal G})}=0\, .
\end{eqnarray}
Thus considering different choices of the Hubble rate $H$, we can get the corresponding solution $F({\cal G})$.

\subsubsection{$\Lambda$CDM Evolution}

Let us firstly realize the $\Lambda$CDM evolution, so consider,
\begin{equation}
P(N)=H_0^2+\frac{1}{6}\rho_0a_0^{-3}e^{-3N}\, .
\end{equation}
where $H_0$ and $\rho_0$ are constants and these correspond to the value of the Hubble rate at present time, and to the cold dark matter respectively. By using Eq. \eqref{eq:RR}, and upon keeping the leading order terms, we may obtain the expression of the $e$-foldings number $N$ in terms of ${\cal G}$ as follows,
\begin{equation}
N=-\frac13\ln \left(\frac{{\cal G}-24 H_0^4}{2\rho_0a_0^-3H_0^2}\right)
\end{equation}
Using the above expression, like in the cases for inflation, one may convert the above equation into a differential equation for the $F({\cal G})$ gravity. The resulting form of the equation is very complicated, however, since we are interested in the low curvature regime, we can keep low order terms in ${\cal G}$. By doing this, we obtain  $F({\cal G})=-c(1- e^{a{\cal G}})$, which is identical to the model I of Eq. \eqref{de} at leading order. In effect, the model I can describe the $\Lambda$CDM successfully, without the need for a cosmological constant term.

\subsubsection{Phantom Evolution}

Let us now demonstrate how to obtain a phantom behavior, without the actual presence of any phantom fluid, so we assume that the function $P(N)$ is \cite{Nojiri:2009kx},
\begin{equation}
P(N)=H_0^2e^{2N/H_0}\, .
\end{equation}
Following the above analysis, the $e$-foldings number $N$ as a function of ${\cal G}$ is equal to,
\begin{equation}
e^{2N/H_0}=\frac{{\cal G}^{1/2}}{24(1+H_0)H_0^3}\, ,
\end{equation}
so by solving the above equation, we get,
\begin{equation}
\frac{4{\cal G}^2}{1+H_0}F''({\cal G})-4{\cal G}F'({\cal G})+F({\cal G})+6{\cal G}=0\, .
\end{equation}
The differential equation above can be solved to yield,
\begin{equation}
F({\cal G})=a {\cal G}^{\alpha}+b{\cal G}^{\beta}
\end{equation}
where $\alpha,\ \beta$ are functions of $H_0$ and $a, \ b$ are integration constants. Thus we see that a phantom dark energy era can also be obtained by some vacuum $F(R, \mathcal{G})$ gravity without the need for a phantom fluid.


\section{Unifying Dark Energy with Inflation in the Context of $F(R,{\cal G})$ Gravity}

Based on the previous sections, we can propose a phenomenological viable model that can describe in a unified way the dark energy and the inflationary epochs, but also that reduces to the standard Einstein-Hilbert description during the matter and radiation domination eras. The functional form of the proposed $F(R,{\cal G})$ gravity is the following,
\begin{equation}
F(R,{\cal G})=R+A{\cal G}^{\alpha}-F_0(1-e^{{\cal{G}}/{\cal G}_0})\, ,
\label{unified}
\end{equation}
and based on the previous sections analysis, the last term in Eq. (\ref{unified}) is dominant only in the low curvature regime, if $\alpha$ is sufficiently larger than unity, for example $\alpha=\mathcal{O}(10)$, so at late times, the $F(R,{\cal G})$ gravity mimics the $\Lambda$CDM model, since we have approximately $F(R,{\cal G})\sim R+F_0$. Also in the high curvature regime, the $F(R,{\cal G})$ gravity is approximately equal to $F(R,{\cal G})\sim \left(\frac{\cal G}{{\cal{G}}_i}\right)^{\alpha}$, where $A=\frac{1}{{\mathcal{G}_i}^{\alpha}}$, hence the inflationary era is dominated by a power law Gauss-Bonnet term. The model of $F(R,{\cal G})$ gravity appearing in Eq. (\ref{unified}) has great similarities with the inflationary exponential $f(R)$ gravity models used in Refs. \cite{Cognola:2007zu,Elizalde:2010ts}, to successfully describe the unification of inflation with the dark energy era. For example, such an exponential $f(R)$ gravity model has the form \cite{Cognola:2007zu,Elizalde:2010ts},
\begin{equation}\label{exponentialfrgravity}
f(R)=R-2\Lambda (1-e^{-\frac{R}{R_0}})+\gamma R^{\alpha}\, .
\end{equation}
Now we shall investigate whether the predicted spectral index of the primordial curvature perturbations for this theory can be compatible with the latest Planck observational data. In order to calculate the spectral index, we shall firstly calculate the evolution of the power spectrum curvature perturbations. Also we shall investigate the behavior of the scalar primordial curvature perturbations, with respect to their evolution after the horizon crossing during the inflationary era.

For the study of the scalar perturbations during the inflationary era, we assumed that the perturbed metric is the following \cite{reviews1},
\begin{eqnarray}
ds^2=-(1+\psi)\, dt^2-2 a (t) \partial_i \beta\, dt\, dx^i+a^2(t) (\delta_{ij}+2\phi \delta_{ij}+2\partial_i \partial_j \gamma)\, dx^i \,dx^j\, ,\label{per}
\end{eqnarray}
where $\psi$, $\beta$, $\gamma$ and $\delta$ represent the smooth perturbation functions. We shall quantify our study of perturbation evolution in $F(R,{\cal G})$ gravity, by using the comoving scalar curvature perturbation, which is defined as follows,
\begin{equation}
\Phi = \phi-\frac{H (\delta F_R+4 H^2\delta F_{\cal G})}{\dot F_R+4 H^2 \dot F_{\cal G}}\, ,
\end{equation}
which is a gauge invariant quantity. It can be shown (see for example \cite{reviews1}), that the differential equation which governs the evolution of scalar perturbations is the following,
\begin{equation}
  \label{eq:ScGen}
  \frac1{a^3\, Q(t)}\partial_t[a^3\, Q(t)\,\dot\Phi]+B_1(t)\, \frac{k^2}{a^2}\,\Phi+B_2(t)\,\frac{k^4}{a^4}\,\Phi=0\, ,
\end{equation}
where  $Q$, $B_1$, and $B_2$ are functions of the $F(R,{\cal G})$ function, of the Hubble rate $H$ and their higher derivatives. We need to note that Eq. (\ref{eq:ScGen}) refers to a general $F(R,\mathcal{G})$ gravity and it has a complicated form and also the last term proportional to $k^4$ is only present for a general $F(R,\mathcal{G})$. As we will show, in the case of an $F(R,\mathcal{G})$ gravity of the form $F(R,\mathcal{G})=R+f(\mathcal{G})$, the above equation is significantly simplified. For the model (\ref{unified}), the master differential equation (\ref{eq:ScGen}) simplifies since $B_2$ is equal to zero, and also $Q(t)$ and $B_1(t)$ are equal to,
\begin{eqnarray}
Q(t)&=&\frac{6 (F_{\cal GG}\dot{\mathcal{G}})^2\left(1
+4F_{\cal GG}\dot{\mathcal{G}}H\right)}{\left(1
+6HF_{\cal GG}\dot{\mathcal{G}}\right)^2} \\
B_1(t)&=&1+\frac{2\dot{H}}{H^2}\, .
\end{eqnarray}
In effect, the master equation (\ref{eq:ScGen}) can be rewritten as follows,
\begin{equation}
a(t)^3Q(t)\ddot{\Phi}+\left(3a(t)^2\dot{a}Q(t)+a(t)^3
\dot{Q}(t)\right)\dot{\Phi}+B_1(t)Q(t)a(t)k^2\Phi=0\, .
\label{diff}
\end{equation}
As we already mentioned, the $F(R,{\cal G})$ gravity during the inflationary era is approximately equal to $\sim A\mathcal{G}^{\alpha}$, which as we showed in previous sections, it generates an approximate power law evolution with scale factor $a(t)\sim t^{\beta}$. Hence, the functions $Q(t)$ and $B_1(t)$ are equal to,
\begin{align}
& Q(t)=-\frac{t^2}{18\beta^2}-\frac{2^{3\alpha}\times 3^{-2+\alpha}A t^4(-1+\alpha)\alpha\left(\frac{(-1+\beta)\beta^3}{t^4}\right)^{\alpha}}{(-1+\beta)\beta^4} \\ \notag &  +\frac{t^2\beta^2 -2t^2\beta^3 +t^2\beta^4}{18\Big{(}\beta^2-\beta^3-2^{3\alpha}\times 3^{\alpha}At^2\alpha\left(\frac{(-1+\beta)\beta^3}{t^4}\right)^{\alpha}+2^{3\alpha}\times 3^{\alpha}At^2\alpha^2\left(\frac{(-1+\beta)\beta^3}{t^4}\right)^{\alpha}\Big{)}^2}\\ \notag &
B_1(t)= 1-\frac{2}{\beta}\, .
\label{QB}
\end{align}
Solving the differential equation (\ref{diff}) analytically is a formidable task, without using an approximation, so we can finding the leading order behavior for small cosmic times, which is,
\begin{equation}
C_0 t^{1+2\beta}\ddot{\Phi}+C_1 t^{2\beta}\dot{\Phi}+C_2 t \Phi=0
\end{equation}
where $C_0$, $C_1$, $C_2$ are equal to,
\begin{eqnarray}
C_0 &=& -\frac{2^{1+3\alpha}\times 3^{-2+\alpha}A\alpha^2(\beta^4)^{\alpha}}{\beta^5}; \nonumber \\
C_1 &=& -\frac{2^{1+3\alpha}\times 3^{-2+\alpha}A\alpha((1+\beta)\beta^3)^{-1\alpha}(4-4\alpha+\beta)}{\beta}; \nonumber \\
C_2 &=& -2 \frac{2^{1+3\alpha}\times 3^{-2+\alpha}A\alpha^2(-2+\beta)(\beta^4)^{\alpha}k^2}{\beta^6}\, .
\label{coefficients}
\end{eqnarray}
Upon solving the above differential equation, we obtain the following solution,
\begin{equation}\label{solbessel}
\Phi(t)= F_1 t^{\mu(1-\beta)}J_{\mu}(mt^{1-\beta})+F_2t^{\mu(1-\beta)}J_{-\mu}(mt^{1-\beta})\, ,
\end{equation}
where the parameters $\mu$, $F_1$ and $F_2$ are given below,
\begin{eqnarray}
\mu &=&\frac{C_0-C_1}{2C_0(1-\beta)};\quad m=\frac{\sqrt{C_2}}{\sqrt{C_0}(1-\beta)}; \\
F_1 &=& C_0^{-\frac{\mu}{2}}C_2^{\frac{\mu}{2}}\left(-2+\frac{2}{\beta}\right)^{-\mu}\beta^{-\mu}P_1 \Gamma(1+\mu); \\
F_2 &=& C_0^{-\frac{\mu}{2}}C_2^{\frac{\mu}{2}}\left(-2+\frac{2}{\beta}\right)^{-\mu}\beta^{-\mu}P_2 \Gamma(1-\mu)\, ,
\label{terms}
\end{eqnarray}
and $P_1$ and $P_2$ are (scale) $k$-dependent constants to be determined later on. By taking the small argument limit of the Bessel function $J_{\mu}(z)$ in the solution (\ref{solbessel}), we have approximately,
\begin{equation}
\Phi(t) \approx F_2 \frac{2^{-\mu}m^{\mu}}{\Gamma(1+\mu)}t^{2\mu(1-\beta)}\, .
\label{pert}
\end{equation}
The power spectrum of the primordial scalar curvature perturbations at horizon crossing is given by,
\begin{equation}
\label{powerspecetrumfgr}
{\cal P}_R=\frac{4\pi
k^3}{(2\pi)^3}\big{|}\Phi\big{|}_{k=aH}^2\, ,
\end{equation}
and we shall reveal the $k$-dependence of each term in the expression of the power spectrum. From Eqs. \eqref{coefficients}, \eqref{terms} and \eqref{pert}, the parameters $C_2$, $F_2$ and  $m$ have a scale dependence, which is,
\begin{equation}
C_2 \propto k^2, \quad F_2 \propto C_2^{\frac{\mu}{2}}, \quad m \propto \sqrt{C_2}\, ,
\end{equation}
therefore we have,
\begin{equation}
{\cal P}_R \propto k^3 |P_2k^{2 \mu}(t-t_s)^{2\mu(1-\beta)}|_{k=aH}^2\, .
\end{equation}
Also the cosmic time can be expressed in terms of the wavenumber $k$, by using the horizon crossing condition $k=aH$, so at leading order we have,
\begin{equation}
t=\left(\frac{k}{\beta}\right)^{\frac{1}{\beta-1}}\, .
\label{time}
\end{equation}
It is important to define the value of the scalar field to have some initial vacuum state value. To this end, we introduce the variable $u=z_s \Phi$, where,
$z_s=Q(t)a(t)$. Then the scalar perturbations action is written in terms of $u$ in the following way,
\begin{equation}
\label{actiaonerenearthebounce}
\mathcal{S}_u\simeq \int d ^3d \tau \left[
\frac{u'}{2}-\frac{1}{2}(\nabla u)^2+\frac{z_s''}{z_s}u^2\right ]\, ,
\end{equation}
where the prime indicates differentiation with respect to the variable $\tau=a^{-1}(t)t$.

So we assume a Bunch-Davies initial state for the vacuum of the scalar field prior to the inflationary era so $u\sim \frac{e^{-ik\tau}}{\sqrt{k}}$. The imaginary phase which depends on the conformal time will be eliminated since eventually we will be interested in the expression $|\Phi (t=t_s)|^2$. By performing the calculation we obtain at the initial time $t_i$ prior to inflation that,
\begin{equation}
\Phi(t_0)=P_2 \propto \frac{1}{k^{\frac{\beta}{\beta-1}}\sqrt{k}Q(t)}\, ,
\end{equation}
so by using an approximate functional form for the function $Q(t)$, we have,
\begin{equation}
Q(t) \approx \frac{t^2}{18\beta(1-\beta)^2}\, ,
\label{Q}
\end{equation}
and eventually, $P_2$ reads,
\begin{equation}\label{akyr1}
P_2 \propto k^{\frac{2}{1-\beta}-\frac{1}{2} -\frac{\beta}{\beta-1}}\, .
\end{equation}
By combining the above resulting behaviors for the parameters, the power spectrum as a function of the wavenumber $k$ is equal to,
\begin{equation}
{\cal P}_R \propto k^{2+\frac{4+2\beta}{1-\beta}}\, ,
\end{equation}
and we can easily extract the spectral index for the primordial scalar curvature perturbations, which is,
\begin{equation}
n_s-1=2+\frac{4+2\beta}{1-\beta}\, .
\end{equation}
Thus, by choosing for example $\beta=177.41$ we obtain approximately $n_s\simeq 0.96$, which is compatible with the 2015 Planck data \cite{Ade:2015lrj}.

Before closing this section, it is worth investigating how the cosmological perturbations evolve after the horizon crossing during inflation, and the differential equation that governs this evolution is \cite{reviews1},
\begin{equation}
\frac{1}{a(t)^3Q(t)}\frac{d }{d t}
\left(a(t)^3Q(t)\dot{\Phi}\right)=0\, ,
\end{equation}
which can be solved to yield,
\begin{equation}
\label{soldiffeqn}
\Phi (t)=D_1+D_2\int
\frac{1}{a(t)^3Q(t)}d t\, ,
\end{equation}
and the function $Q(t)$ is given in Eq. \eqref{Q} and $a(t)\sim t^{\beta}$. Thus, by evaluating the second term in the above equation we get approximately,
\begin{equation}
\int \frac{1}{a(t)^3Q(t)}dt \propto t^{-1-\beta}\, .
\end{equation}
Hence, as the cosmic time increases, the perturbations after the horizon crossing remain approximately constant, which is a good feature, since this will render them relevant to present day observations, after they reenter the horizon during the radiation domination era.

In conclusion, the proposed model in Eq. (\ref{unified}), apart from providing a theoretical unified framework which can qualitatively describe inflation and the dark energy era, it also provides a quantitatively viable inflationary era, compatible with the observations.

\section{Conclusions}

In this paper we studied certain classes of phenomenologically appealing $F(R,{\cal G})$ gravities. We focused on the successful realization of the inflationary and dark energy eras, but also we investigated how the intermediate matter and radiation domination eras can be realized. As we discussed, the most appealing class of $F({\cal G})$ is types of theories which have the form $F(R,{\cal G})=R+f(\mathcal{G})$, since for these no superluminal modes appear in the evolution of scalar perturbations, which would be proportional to $k^4$, only $k^2$ modes appear. In our study we presented how various inflationary and dark energy scenarios can be realized with this type of $F(R,{\cal G})$ gravity, and also we investigated how a quasi-de Sitter evolution can be realized with a more general form of $F(R,{\cal G})$. Finally, we proposed a unification model, which at early times produces a nearly scale invariant power spectrum of primordial curvature perturbations, at intermediate times, and particularly, during the matter and radiation domination eras, it behaves like ordinary Einstein-Hilbert gravity, and at late-time it produces a dark energy era. This model should be accompanied by two ordinary perfect fluids, describing the matter and radiation domination eras, and these produce insignificant effects at early and late times. The study of more general $F(R,{\cal G})$ gravities in detail is a demanding task, and to some extend problematic, due to the fact that the calculation of the scalar primordial curvature perturbations is highly non-trivial and too complicated, plus superluminal modes proportional to $k^4$ powers of the wavenumber appear. However simpler models of $F(R,\mathcal{G})$ gravity like the one appearing in Eq. (\ref{unified}) can provide a fertile ground for cosmological phenomenology, which is also free from $k^4$ instabilities in the perturbations.

\section*{Acknowledgments}

This work is supported by MINECO (Spain), FIS2016-76363-P (S.D.O), by project 2017 SGR247 (AGAUR, Catalonia) (S.D. Odintsov) and by PHAROS-COST action No: CA16214 (S.D. Odintsov and V.K. Oikonomou).

\section*{Appendix: Explicit Form of Parameters}

Here we quote the explicit form of the parameters $\Gamma_i$, $i=1,...,4$ and $\gamma$ appearing in Eq. (\ref{frgfinal}) in the text. These are,
\begin{align}\label{appendixprmters}
& \Gamma_1=-96 \beta  H_0^2 H_i \left(H_0 \left(\sqrt{H_0^2+8 H_i}-4 H_0^2+H_0\right)+8 H_i^2\right)\, ,\\ \notag &
\Gamma_2=12 c_2 H_0 H_i-\frac{144 \beta  H_0^3 H_i \left(\sqrt{H_0^2+8 H_i}+H_0\right) \left(-2 H_0^3+\frac{1}{2} H_0 \left(\sqrt{H_0^2+8 H_i}+H_0\right)+4 H_i^2\right)}{\frac{1}{4} \left(\sqrt{H_0^2+8 H_i}+H_0\right)^2-\frac{1}{2} H_0 \left(\sqrt{H_0^2+8 H_i}+H_0\right)-2 H_i}\, , \\ \notag &
\Gamma_3=\frac{24 \beta  H_0 H_i}{H_0+\mu },\,\,\, \Gamma_4=288 \beta  H_0^4 H_i\, , \\ \notag &
\gamma=3 \mathcal{C}_2 H_0^2+3 \mathcal{C}_2 H_0 \mu -6 \mathcal{C}_2 H_i+12 \beta  H_0^4+12 \beta  H_0^3 \mu -72 \beta  H_0^2 H_i\, .
\end{align}

\end{document}